\documentstyle[11pt,aaspp4]{article}

\def\mic{~$\mu$m}
\def\kms{~km s$^{-1}$}
\def\gtrapprox{\;\lower 0.5ex\hbox{$\buildrel >
    \over \sim\ $}}		
\def\lessapprox{\;\lower 0.5ex\hbox{$\buildrel < \over \sim\ $}}

\begin{document}

\title{Detection of CO($3-2$) Emission at $z = 2.64$ from the Gravitationally 
Lensed Quasar MG 0414+0534}

\author{Richard Barvainis }
\affil{MIT Haystack Observatory, Westford, MA 01886 USA\footnote
{Radio Astronomy at the Haystack Observatory of the
Northeast Radio Observatory Corporation (NEROC) is supported by a grant
from the National Science Foundation}}

\author{Danielle Alloin}
\affil{Service d'Astrophysique, L'Orme des Merisiers,
CE Saclay, 91191, Gif-Sur-Yvette, Cedex, France}

\author{Stephane Guilloteau}
\affil{IRAM, DomaineUniversitaire de Grenoble, F-38460, St. Martin d'H\`eres,
France}

\author{Robert Antonucci}
\affil{UC Santa Barbara, Physics Department, Santa Barbara, CA 93106 USA}

\begin{abstract}
We have detected CO($3-2$) line emission from the gravitationally
lensed quasar MG 0414+0534 at redshift 2.64, using the IRAM Plateau
de Bure Interferometer.  The line is broad, with 
$\Delta v_{\rm FWHM} = 
580$\kms .   The velocity-integrated CO flux is comparable to, 
but somewhat smaller than, that of IRAS F10214+4724 and the Cloverleaf 
quasar (H1413+117), both of which are at similar redshifts.  The 
lensed components 
A1+A2 and B were resolved, and separate spectra are presented for each. 

We also observed the unlensed radio quiet quasar PG 1634+706 at $z=1.33$,  
finding no significant CO emission.   

\end{abstract}


\section{INTRODUCTION}

The study of molecular gas in active galactic nuclei (AGNs) can provide
a window on the processes that fuel, and sometimes hide, the central
engine, and can potentially aid in understanding how large scale
processes such as galaxy interactions and mergers trigger nuclear
activity.  The general picture of AGN activation that has emerged over
the past decade or so involves the following events:  a merger induces a
concentration of mass in the galactic nucleus, in the form of dusty
molecular gas (e.g.  Barnes \& Hernquist 1996).  Some fraction of this
molecular material is driven down to very small scales, where it
provides the fuel for the central black hole.  Initially, the newly
energized AGN is hidden by the dusty gas, but it eventually emerges (at
least along some lines of sight) as the gas is consumed, ejected, or
flattened into a disk-like configuration.  Along the way, a nuclear
starburst may occur.  Absorption and re-emission of the nuclear and
stellar energy sources by dust can result, at least in the early phases,
in a very luminous infrared source.  This sort of scenario provides a
natural evolutionary link between luminous infrared galaxies and mature
AGNs (Sanders et al 1988).

 Carbon monoxide and a limited suite of other molecules have been
detected, and in few cases mapped, in a number of AGNs.  At low
redshifts these include Seyfert galaxies (Sanders et al 1987, 1989;
Heckman et al 1989; Miexner et al 1990; Sternberg, Genzel \& Tacconi
1994; Helfer \& Blitz 1995), radio galaxies (Mirabel, Sanders, \&
Kaz\`es 1989), and low- to moderate-luminosity quasars (Sanders,
Scoville, \& Soifer 1988; Barvainis, Alloin, \& Antonucci 1989; Alloin
et al 1992; Scoville et al 1993).  At high redshifts ($z > 0.4$) there
have been only three confirmed CO detections, amidst a number of (as
yet) unconfirmed reports.
The confirmed high-$z$ CO sources are the
infrared galaxy/hidden quasar IRAS F10214+4724 at $z = 2.28$ (Brown \&
Vanden Bout 1992; Solomon et al 1992), the Cloverleaf quasar (H1413+117)
at $z = 2.56$ (Barvainis et al 1994; Wilner, Zhao, \& Ho 1995), and the
quasar BR1202$-$0725 at $z = 4.69$ (Omont et al 1996a; Ohta et al 1996).
It appears from these high-redshift objects, and also from molecular studies at
lower redshifts, that a necessary but not sufficient condition for CO
detection is the presense of a strong far-infrared or
millimeter/submillimeter continuum, i.e.  one that is detectable either
in the IRAS database at 60\mic\ or 100\mic , or by ground-based
telescopes at $\sim \lambda 1$mm.

  Of these three confirmed high-$z$ CO sources, the Cloverleaf (4
images; Magain et al 1988) and IRAS F10214+4724 (gravitational arc
system; e.g.\ Broadhurst \& Lehar 1996; Eisenhardt et al 1996) are
definitely gravitationally lensed, and both have IRAS and 1 mm continuum
detections (Barvainis et al 1995, and references therein).
BR1202$-$0725 may also be lensed, but this is uncertain at this time.
The optical emission shows a single source (a classical broad-line
quasar), but in the 1 mm continuum and in CO($5-4$) line emission the
source is a double with a separation of $\approx 4''$.  The southeast
mm/CO source is coincident with the optical quasar, while the location
of the northwest component shows no counterpart in deep optical
exposures (Hu, McMahon, \& Egami 1996).  The CO spectra of the two
components are rather noisy, but reveal no significant differences in
line shape or velocity centroid.  No direct evidence has yet been found
for a lensing galaxy or cluster toward BR1202$-$0725, but evidence of
gravitational shear in the field has been reported (see Omont et al
1996a).  Either the system is lensed, with the optical part of the
northwest image extincted by dust in the lensing galaxy, or else it
consists of a luminous quasar with extremely luminous submillimeter and
CO emission and an optically weak or absorbed (but equally mm/CO strong)
companion.

   It would be very useful to know to what extent gravitational lensing
amplification is in fact necessary for the detection and study of CO and
other molecules at high redshift, given current instrumental
limitations.  Recent work suggests that lensing, while always helpful,
may not be required in all cases.  In the apparently unlensed radio
galaxy 53W002 ($z=2.39$), a weak CO(3--2) line has recently been
reported by Scoville et al (1997).  This object has not been detected as
yet in the far-infrared/submillimeter continuum.  Guilloteau et al
(1997) report a highly significant CO(5--4) line at $z = 4.41$ from the
radio quiet quasar BRI 1335-0415, which has a 1.3 mm continuum 
detection (Omont et al 1996b) but like 53W002 shows no evidence for
lensing.  In this Letter we report a sensitive search for CO emission
from two high redshift systems, both of which are far-infrared/submillimeter
continuum sources, but only one of which is gravitationally lensed.

\section{OBSERVATIONS AND RESULTS}

The observed sources are PG 1634+706, a luminous radio-quiet quasar at
$z=1.3371$, and MG 0414+0534 (= 4C+05.19), a gravitationally lensed
quasar at $z = 2.639$.

  The observations were carried out in the $\lambda$3mm band using the
Plateau de Bure Interferometer of the Institut de Radio Astronomie
Millimetrique over a number of sessions during 1996 and 1997.  The array
consisted of 4 or $5\times 15$m antennas in various configurations
covering baselines from 32--400 m.  Total observing times were 10 hours
for PG 1634+706 and 30 hours for MG 0414+0534.  The total frequency
bandwidth was 500 MHz, and system temperatures were typically 130 K.
For MG 0414+0534, which has 40 mJy of continuum flux, bandpass
correction was applied using 3C454.3, 0415+379, and 0420-014 as the
bandpass calibrators.  We estimate that the bandpass correction is
accurate to better than a few percent, or $\lesssim 1$ mJy, across the
observed band.

  Significant CO emission was detected only in MG 0414+0534.  Line 
parameters for MG 0414+0534, and upper limits for PG 1634+706,
are summarized in Table 1.

\subsection{PG 1634+706}

  PG 1634+706 was detected in all four IRAS bands (12, 25, 60, \&
100\mic ) in pointed observations (Neugebauer et al 1986).  Given its
radio quiet nature, the mid- to far-infrared continuum is very likely emission
from dust (see, e.g., Barvainis 1992).  The IRAS fluxes are similar to
those of the Cloverleaf at 60 and 100\mic , so PG 1634+706 seemed like
an excellent candidate for study in molecular line emission.  Because of
the narrow bandwidths available in millimeter-wavelength spectrometers,
it is important to use the best possible estimate of the systemic
redshift of the target.  The standard catalog redshift of PG 1634+706 is
1.334, based on the high-ionization broad emission lines.  However,
these lines can be blue-shifted relative to systemic by over 1000\kms\
in some quasars.  We used instead the (adjusted) systemic redshift
estimate given by Tytler and Fan (1992) of $z= 1.3371$, to obtain an
observing frequency of 98.647 GHz for the CO($2-1$) line (rest frequency
230.538 GHz).

  No significant line emission was detected toward PG 1634+706.  It is
very unlikely that the source was resolved out, since the synthesized
beam was $5.7\times 5.2''$ (PA $-27^{\circ}$), or $\sim 30h_{70}^{-1}$ kpc in
diameter at the source (where $h_{70}$ is the Hubble constant in units of 70 
km s$^{-1}$ Mpc$^{-1}$, and we have taken $q_0 = 1/2$).  The spectrum
is shown in Figure 1, smoothed to a velocity resolution of 152\kms .
The RMS noise per channel at this resolution is 1.3 mJy/beam; using this
we derive a $3\sigma$ upper limit to the line peak of 2.9 mJy and to the
velocity-integrated line flux of 0.86 Jy km s$^{-1}$, assuming a
gaussian line of full width half maximum 300\kms .  There is a weak
continuum source, with flux density $1.3\pm 0.4$ mJy, at the position of
the quasar.  This is comparable to the $\sim 1$ mJy flat spectrum
centimeter wavelength source (Antonucci \& Barvainis 1988), showing that
this component remains flat out to millimeter wavelengths.

\subsection{MG 0414+0534}

  MG 0414+0534 resolves into four components with maximum separation
$2.1''$ in high resolution radio images (A1, A2, B, and C; Hewitt et al
1992).  The combined optical spectrum between 6000\AA\ and 9400\AA\ is
extremely red, with $S_{\nu} \propto \nu^{-8.8}$ (Hewitt et al 1992).
The lensed nature of this system was confirmed by optical spectra
comparing components A1+A2 and B (Angonin-Willaime et al 1994).
Lawrence et al (1995, herafter LEJT) identified a strong broad H$\alpha$
line from near-infrared spectroscopy and determined a redshift of $z = 2.639
\pm 0.002$ for the lensed object; broad H$\alpha$ is known to be a good
indicator of the true systemic redshift.  The redshift of the lensing
galaxy remains unknown.  LEJT argue that MG 0414+0534 is a typical
high-redshift quasar heavily reddened by dust in the lens.  They argue
further that the IRAS detections at 25\mic\ ($70\pm 24$ mJy) and 60\mic\
($140\pm 40$ mJy) originate in the quasar itself, rather than in the
lens, for the following reasons.  First, the overall dereddened SED,
including the IRAS points, is typical of a low-redshift IRAS quasar.
Second, if the lens is a normal galaxy at an intermediate redshift
(0.5--1.5), it would have to have an unprecedented infrared luminosity
to be detected by IRAS.  MG 0414+0534 is thus very similar to the
Cloverleaf:  a distant, lensed quasar with strong infrared emission.  We
searched for CO($3-2$) emission at a frequency of 95.025 GHz, based on
the H$\alpha$ redshift of LEJT.

 MG 0414+0534 showed evidence for CO emission after the first 8 hours of
observation.  The line was confirmed in 22 hours of follow-up, resulting
in a secure detection of CO($3-2$) emission centered within 100\kms\ of
the H$\alpha$ redshift.  The final spectrum, spatially integrated over
the four centimeter-wavelength continuum components, is shown in Figure
2 (top panel).

The lensed configuration of the centimeter radio continuum components is
as follows:  relative to component A1 in the 5 GHz map of Hewitt et al
(1992), A2 lies at [RA, Dec] = [$+0.13'',+0.39''$], B lies at
[$-0.60'',+1.89''$], and C lies at [$-1.94'',+0.25''$].  The relative
strengths, combining A1+A2 into one component designated A, are
approximately 14:3:1 for A:B:C.  With a synthesized beam of $2.0 \times
0.9''$ (PA $16\arcdeg$), it is difficult to separate the components in
the CO($3-2$) velocity-integrated map plane.  However, by fitting the UV
data directly we were able to resolve the combined A components from
component B and obtain separate CO($3-2$) spectra.  These are shown in
Figure 2 (two lower panels).  The line from B is weak, and looks
somewhat different from the A line.  The relative strength A:B in the
continuum at 5 GHz is 5:1.  In our millimeter wavelength data the
continuum ratio is more like 7:1 (see Figure 3).  Given the noise on the
spectra, and assuming that the 7:1 ratio holds for CO, the B line-shape
is only marginally consistent with that of A.  The only channel in the B
spectrum that is very significantly different from the prediction based
on A/7 is the one centered at +63\kms , which is $3.9\sigma$ too high.
But the channels centered at +221\kms\ and $-95$\kms\ are also too high,
by $1.4\sigma$ and $2.1\sigma$, respectively.  To explain a difference
between the A and B spectra, it is quite possible that due to proximity
to a caustic the B and A images sample a different region in an extended
CO source, resulting in a different line shape.  Further progress on
this point will have to await data of higher resolution and
signal-to-noise ratio.

There is some evidence that the CO source may in fact be slightly 
extended and displaced about $1''$ west of the continuum source, but the 
size measurement is complicated by the relatively strong (40 mJy)
continuum source underlying the CO line emission.
Until more sensitive observations are performed,
we quote a limit on the CO source size of $\theta_{\rm CO} < 3''$.

\section{Discussion and Conclusions}

The detection of CO in MG 0414+0534 adds to the similarities between it
and both IRAS F10214+4724 and the Cloverleaf -- all have weak IRAS
detections, are gravitationally lensed, lie at redshifts near 2.5, and
have abundant molecular gas.  Given that detectable mid- to far-infrared
flux appears to be a good (but not infallible) selection criterion for
molecular gas, it is useful to look at the CO-to-IR luminosity ratio.
We define the frequency-integrated CO($3-2$) luminosity relative to the
$\nu L_{\nu}$ luminosity at 30\mic\ (rest frame, or $\approx 100$\mic\
observed frame for $z\approx 2.5$) to be $R_{\rm CO/IR} = \int
L_{\nu_e}d\nu_e[{\rm CO(3-2)}]/\nu_e L_{\nu_e}[30~\mu{\rm m}]$.  For MG
0414+0534, $R_{\rm CO/IR} = 1.0\times 10^{-6}$, which is equal to the
value found for F10214+4724 and 2.4 times lower than that of the
Cloverleaf.

For a sample of seven nearby quasars and luminous Seyfert 1
galaxies\footnote{3C48, MRK 1014, I Zw 1, MRK 876, MRK 817, MRK 231, and
IRAS 07598+6508.}  detected in CO($1-0$), this ratio spans the range
$R_{\rm CO/IR} = (4.7-15\times 10^{-6}) {T_B(3-2)\over T_B(1-0)}$.  Here
we have included a factor for the line brightness temperature ratio
between CO(3--2) and CO(1--0).  This ratio may be $\sim 0.5$--0.8 due to
subthermal excitation or other effects, as observed in luminous
infrared galaxies at low redshift (Radford, Solomon, \& Downes 1991).
For the detected high redshift quasars $R_{\rm CO/IR} = 1.0-2.4\times
10^{-6}$, a factor of five below the local objects if the CO($3-2$) and
CO($1-0$) brightness temperatures are equal.  It is likely however that
${T_B(3-2)\over T_B(1-0)} < 1$, as noted, bringing the CO-to-IR ratios
closer.  Also, the nearby sample only includes CO-detected objects,
which would tend to have a higher ratio than average (to our knowledge
no complete CO survey of nearby IRAS-detected objects has been done).
Parenthetically, the CO-to-IR ratio could be affected by lensing:  if
the CO and infrared sources are not the same size, they might be
magnified by different factors.

Because of its IRAS detections, PG 1634+706 seemed like a good candidate
for CO emission.  For this quasar $R_{\rm CO/IR} < (0.7\times
10^{-6}){T_B(3-2)\over T_B(2-1)}$, which is below the lowest values for
the detected objects noted above.  Since PG 1634+706 is the only
unlensed radio quiet quasar detected by IRAS at high redshift, it is
anomalously infrared-luminous, and this may be related to the low CO-to-IR
ratio.  Sanders, Scoville, \& Soifer (1991) note that among nearby
luminous IRAS galaxies, the highest luminosity galaxies have the lowest
CO-to-IR ratios (or, in their formulation, the highest $L_{\rm
IR}/M(H_2)$ ratios).  This could be a result of the extra energy input
from an active galactic nucleus heating the circumnuclear dust.  PG
1634+706 is a very powerful quasar whose UV emission may drive up dust
temperatures over a very large volume.

Using the CO size upper limit of $3''$, the upper limit on the dynamical mass in
MG 0414+0523 is 
\begin{equation}
M_{\rm dyn}\approx {r \Delta V_{\rm FWHM}^2\over G \sin^2 i} <
9.0\times 10^{11} m^{-1} h_{70}^{-1}\  M_{\sun},
\end{equation}
where $m$ is the (unknown) lensing magnification factor, and the 
inclination angle $i$ has been taken to be $45^{\circ}$.
A very conservative upper limit to the molecular hydrogen mass is 
$M({\rm H}_2) < 5\times L'_{\rm CO} = 2.2\times 10^{11} m^{-1} 
h_{70}^{-2}\  M_{\sun}$
(this would correspond to Galactic molecular clouds, whereas in infrared
galaxies and quasars it is thought that the  
H$_2$ mass to CO luminosity ratio is much lower -- see Barvainis et al 1997;
Shier, Rieke, \& Reike 1994).
So the molecular hydrogen mass is well below the current limit on the dynamical
mass.  A value or a better limit will follow with high resolution measurements of 
CO($7-6$) at 1 mm wavelength.
 
Finally, it is worth noting that CO fluxes are not subject to
microlensing, reddening, or intrinsic variability like the optical
fluxes are.  When sensitive, high resolution ($\approx 0.5''$) CO
observations become available they can be used as an accurate data set
for helping constrain the macrolensing parameters of the gravitational
lens system toward MG 0414+0534, as has recently been done for the
Cloverleaf quasar (Kneib et al 1997).

\acknowledgments

\begin{deluxetable}{lccc}
\scriptsize
\tablecaption{Line Parameters \label{tbl-1}}
\tablehead{
\colhead{} & \colhead{PG 1634+706} & 
\colhead{MG 0414+0534}& \colhead{Units}
}
\startdata
Line................. & CO($2-1$) & CO($3-2$) &... \nl
$z_{\rm obs}$.................&1.337&  2.639 &...   \nl
$z_{\rm CO}$.................  &... & 2.639 &...  \nl
$S_{\rm CO}$(peak)...... &$<3.7$  &4.4 &mJy  \nl
$\Delta v_{\rm FWHM}$........ &... &580 &\kms\ \nl
$S_{\rm CO}\Delta v$........... &$<0.9$& 2.6 & Jy\kms\ \nl
$L_{\rm CO}$................ &$<4.4$ 
&$57 m^{-1}$ &$10^6 h_{70}^{-2}\ L_{\sun}$      \nl
$L'_{\rm CO}$................ &$<1.0 $ 
&$4.3 m^{-1} $ &$10^{10}h_{70}^{-2}$\ K\kms\ pc$^2$ \nl
\enddata
\tablecomments{The variable $m$ in the luminosities for MG 0414+0534 is
the (unknown) gravitational lensing magnification factor.}
\end{deluxetable}

\vglue 1.0truecm
\centerline {\bf FIGURE CAPTIONS}

\figcaption{CO($2-1$) spectrum toward PG 1634+706, smoothed to a resolution of 
152\kms .  The velocity offsets are relative to a redshift of 1.337, and the
horizontal dashed line represents the continuum level.
 \label{fig1}}

\figcaption{CO($3-2$) spectra toward MG 0414+0534, smoothed to a resolution of 
158\kms .  The velocity offsets are relative to a redshift of 2.639.
The lower panel shows the spectrum of lensed component A (= A1+A2), 
the middle panel component B, and the top panel the sum.
 \label{fig2}}

\begin{thebibliography}{}

\bibitem[]{} Alloin, D., Barvainis, R., Gordon, M.A., \& Antonucci, R.R.J.\
1992, A\&A, 265, 429

\bibitem[]{} Angonin-Willaime, M.C., Vanderriest, C., Hammer, F.,
\& Magain, P.\ 1994, A\&A, 281, 388

\bibitem[]{} Antonucci, R., \& Barvainis, R.\ 1988, ApJ, 332, L13

\bibitem[]{} Barnes, J.E., \& Hernquist, L.\ 1996, ApJ, 471, 115
 
\bibitem[]{}
Barvainis, R., Alloin, D., \& Antonucci, R.\ 1989, ApJ, 337, L69
 
\bibitem[]{}
Barvainis, R.\ 1992, In {\it Testing the AGN Paradigm}, ed. S.S. Holt,
S.G. Neff, \& C.M. Urry, p. 129.  New York: AIP Conference Proceedings
 
\bibitem[]{}
Barvainis, R., Tacconi, L., Alloin, D., Antonucci, R., \& Coleman, P.\ 
1994, Nature, 371, 586
 
\bibitem[]{}
Barvainis, R., Antonucci, R., Hurt, T., Coleman, P., \& Reuter, H.-P.\ 1995, 
ApJ, 451, L9

\bibitem[]{} Barvainis, R., Maloney, P., Antonucci, R., \& Alloin, D.\ 1997,
ApJ, 484, 695

\bibitem[]{} Broadhurst, T., \& Lehar, J.\ 1995, ApJ, 450, L41

\bibitem[]{}
Brown, R.L., \& Vanden Bout, P.A.\ 1992, ApJ, 397, L19            

\bibitem[]{} Eisenhardt, P.R., Armus, L., Hogg, D.W., Soifer, B.T., 
Neugebauer, G., \& WErner, M.W.\ 1996, ApJ, 461, 72

\bibitem[]{} Guilloteau, S., Omont, A., McMahon, R.G., Cox, P., \& Petitjean, P.\ 1997,
A\&A, in press

\bibitem[]{} Heckman, T.M., Blitz, L., Wilson, A.S., Armux, L., \& Miley, G.k.\
1989, ApJ, 342, 735

\bibitem[]{} Helfer, T., \& Blitz, L.\ 1995, ApJ, 450, 90

\bibitem[]{} Hewitt, J.N., Turner, E.L., Lawrence, C.R., Schneider, D.P.,
\& Brody, J.P.\ 1992, AJ, 104, 968

\bibitem[]{} Hu, E.M., McMahon, R.G., \& Egami, E.\ ApJ, 459, L53

\bibitem[]{} Kneib, J.-P., Alloin, D., Mellier, Y., Guilloteau, S., Barvainis,
R., \& Antonucci, R.\ 1997, A\&A, in press 

\bibitem[]{} Lawrence, C.R., Elston, R., Januzzi, B.T., \& Turner, E.L.\ 1995,
AJ, 110, 2570 (LEJT)

\bibitem[]{} Magain, P., et al 1988, Nature, 334, 325

\bibitem[]{} Miexner, M., Puchalsky, R., Blitz, L., Wright, M.,
\& Heckman, T.\ 1990, ApJ, 354, 158

\bibitem[]{} Mirabel, O.F., Sanders, D.B., \& Kaz\'es, I.\ 1989, ApJ, 340, L9
 
\bibitem[]{} Neugebauer, G., Miley, G.K., Soifer, B.T., \& Clegg, P.E.\ 1986,
ApJ, 308, 815

\bibitem[]{} Ohta, K., Yamada, T., Nakanishi, K., Kohno, K.,
Akiyama, M., \& Kawabe, R.\ 1996, Nature, 382, 426

\bibitem[]{} Omont, A., Petijean, P., Guilloteau, S., McMahanon, R.G.,
Solomon, P.M., \& Pecontal, E.\ 1996a, Nature, 382, 428

\bibitem[]{} Omont, A., McMahon, R.G., Cox, P., Kreysa, E., Bergeron, J., Pajot, F.,
\& Storrie-Lombardi, L.J.\ 1996b, A\&A, 315, 1

\bibitem[]{} Radford, S.J.E., Solomon, P.M., \& Downes, D.\ 1991, ApJ, 368, L15

\bibitem[]{} Sanders, D.B., et al 1988, ApJ, 325, 74

\bibitem[]{} Sanders, D.B., Young, J.S., Scoville, N.Z., Soifer, B.T., \&
Danielson, G.E.\ 1987, ApJ, 312, L5

\bibitem[]{} Sanders, D.B., Scoville, N.Z., \& Soifer, B.T.\ 1988, ApJ, 335, L1 

\bibitem[]{} Sanders, D.B., Scoville, N.Z., Zensus, A., Soifer, B.T., Wilson,
T.L., Zylka, R., \& Steppe, H.\ 1989, A\&A, 213, L5

\bibitem[]{} Sanders, D.B., Scoville, N.Z., \& Soifer, B.T.\ 1991, ApJ, 370, 158

\bibitem[]{} Scoville, N.Z., Padin, S., Sanders, D.B., Soifer, B.T., \&
Yun, M.S.\ 1993, ApJ, 415, L75

\bibitem[]{} Scoville, N.Z., Yun, M.S., Windhorst, R.A., Keel, W.C., 
\& Armus, L.\ 1997, ApJ, 485, L21

\bibitem[]{} Shier, L.M., Rieke, M.J., \& Reike, G.H.\ 1994, ApJ, 433, L9

\bibitem[]{} Solomon, P.M., Downes, D., \& Radford, S.J.E.\ 1992, ApJ, 398, L29

\bibitem[]{} Sternberg, A., Genzel, R., \& Tacconi, L.\ 1994, ApJ, 436, L131 

\bibitem[]{}  Tytler, D., \& Fan, X.\ 1992, ApJSupp, 79, 1

\bibitem[]{} Wilner, D.J., Zhao, J.-H., \& Ho, P.T.P.\ 1995, ApJ, 452, L91


%
\end{thebibliography}
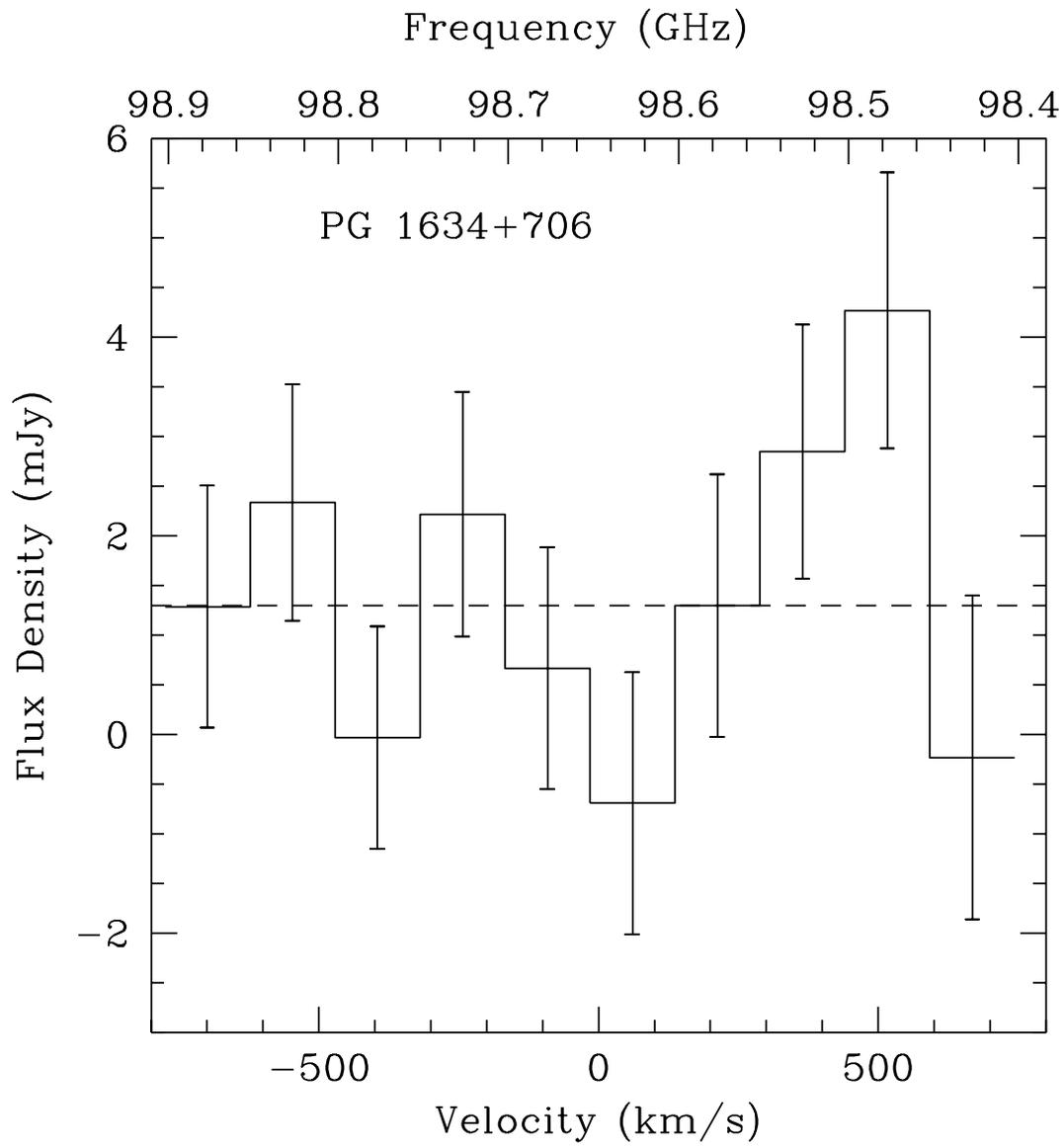
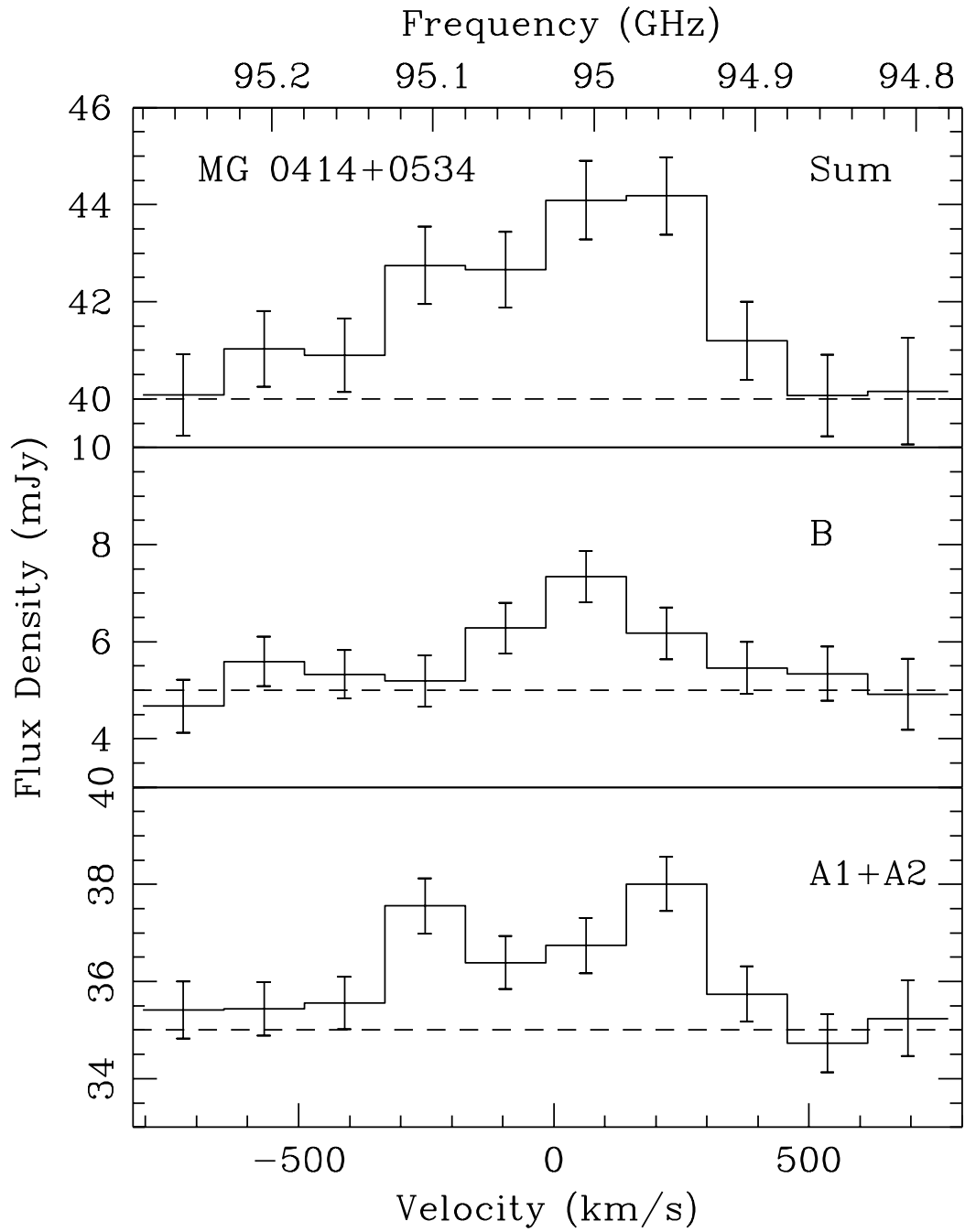
\end{document}